# Properties of polycrystalline nanoparticles with uniaxial and cubic types of magnetic anisotropy of individual grains

V. A. Bautin[1], A. G. Seferyan[1], M. S. Nesmeyanov[3] and N. A. Usov[1,2,3]

[1]*National University of Science and Technology «MISiS», 119049, Moscow, Russia*
[2]*Pushkov Institute of Terrestrial Magnetism, Ionosphere and Radio Wave Propagation, Russian Academy of Sciences, IZMIRAN, 142190, Troitsk, Moscow, Russia*
[3]*National Research Nuclear University "MEPhI", 115409, Moscow, Russia*

**Abstract**

The influence of the crystal structure inhomogeneities on the magnetic properties of cobalt nanoparticles with different aspect ratio and spherical nanoparticles of chromium dioxide, cobalt ferrite and magnetite has been studied by means of numerical simulation. The polycrystalline nanoparticles are modeled by means of subdivision of the nanoparticle volume into tightly bound single-crystal granules with randomly distributed directions of the easy anisotropy axes. The probability of appearance of quasi uniform and vortex states in sufficiently large assemblies of polycrystalline nanoparticles of various types have been calculated depending on the nanoparticle diameter. It is shown that the subdivision of a nanoparticle into single-crystal granules with different orientations of the easy anisotropy axes substantially reduces the effective single-domain diameters for particles with uniaxial type of anisotropy of individual granules. However, for particles with cubic type of magnetic anisotropy the influence of the crystal structure inhomogeneities on the equilibrium properties of the particles is not so important even for magnetically hard cobalt ferrite nanoparticles. It is practically absent for magnetically soft magnetite nanoparticles.

**PACS** 75.20.-g; 75.50.Tt; 75.40.Mg

**Keywords:** Polycrystalline magnetic nanoparticles, Magnetic anisotropy, Single-domain diameter, Numerical simulation

## I. INTRODUCTION

Magnetic nanoparticle assemblies are promising for various technological and biomedical applications, in particular for magnetic resonance imaging contrast enhancement, targeted drug delivery and magnetic hyperthermia [1-3]. However, in order to fully understand the magnetic properties of nanoparticles, detailed experimental information on the particle crystal structure is important. Are the magnetic particles single-crystal or do they consist of a set of several monocrystalline grains? In the latter case what the average size of the monocrystalline granules is? What are the type of magnetic anisotropy and the direction of the easy anisotropy axes of individual granules in this nanoparticle? Such detailed information is available in some cases for very small nanoparticles with dimensions on the order of several nanometers [4,5]. But for magnetic nanoparticles of larger sizes experimental information on the particle crystal structure is scarce. It is usually tacitly assumed that magnetic nanoparticles with dimensions on the order of several tens of nanometers, which are used in particular in biomedical applications [2,3], are monocrystalline. However, it does remain unclear why the properties of these nanoparticles, for example, the saturation magnetization, the magnetic anisotropy constants, etc. differ substantially from the corresponding bulk values.

Meanwhile, the formation of polycrystalline particles in the course of chemical reactions is quite likely [6], if the growth of large nanoparticles occurs as a result of the coalescence of single-crystal nuclei. It has been found recently, that monodisperse iron oxide nanoparticles of sufficiently large sizes prepared by thermal decomposition of organometallic precursors [7] or other methods [8-11] may have complicated crystalline structure. Multi-core nanoparticles [12,13] consisting of monocrystalline magnetic grains can also be considered as polycrystalline nanoparticles if there is exchange interaction of appreciable value between the constituting grains.

It should be noted that now, due to a significant increase in computer performance, numerical modeling makes it possible to study the equilibrium and kinetic properties of magnetic nanoparticles in all details. Single-domain diameters of different types of magnetic particles can be obtained both with the help of analytical estimates [14-18] and using numerical modeling [18-24]. It was shown in particular [23] that effective single-domain diameters of non ellipsoidal nanoparticles, such as cube or cylinder, do not differ considerably from single-domain diameter of an ellipsoid of a proper shape. At the same time, the experimental values for single-domain diameters of various magnetic nanoparticles, their dependence on the particle shape, possible composition variations, crystal structure, etc., are practically unknown.

Investigation of the effect of the crystal structure heterogeneities on the equilibrium magnetic properties of nanoparticles is important also for understanding the behavior of nanoparticles in a quasistatic and alternating magnetic field. Recently [25], the magnetic properties of polycrystalline cobalt nanoparticles of spherical shape have been studied theoretically. In this paper, single-domain diameters for oblate and elongated single-crystal cobalt particles, as well as for spherical nanoparticles of chromium dioxide, cobalt ferrite, and magnetite are determined using numerical simulation. Particles of cobalt and chromium dioxide have uniaxial type of magnetic anisotropy, whereas particles of cobalt ferrite and magnetite belong to the cubic type of anisotropy, with a different number of equivalent directions of the easy anisotropy axes. In addition, the nanoparticles studied have different magnetic hardness. The latter is characterized by a dimensionless parameter $p = N_z M_s^2 / 2K$ [14,17,18], where $N_z$ is the demagnetizing factor of the spheroidal particle along the symmetry axis, $M_s$ is the saturation magnetization, and $K$ is the absolute value of the particle anisotropy constant.

Analogous calculations have been also carried out for assemblies of polycrystalline nanoparticles of the same compositions, with different amounts of single-crystal granules in the particle volume. In the absence of more specific information, it is assumed that the easy anisotropy axes in various single-crystal granules of a polycrystalline nanoparticle are randomly oriented. It is shown that for polycrystalline nanoparticles with uniaxial anisotropy the effective single-domain diameter $D_{c,ef}$ is significantly reduced in comparison with that of monocrystalline nanoparticle, $D_{c0}$. The difference between these diameters increases with increasing magnetic hardness of the nanoparticle, that is, with a decrease in the parameter $p$. On the other hand, for nanoparticles with a cubic type of magnetic anisotropy, a decrease in the effective single-domain diameter was obtained only for magnetically hard nanoparticles of cobalt ferrite. It has been found that the diameters $D_{c,ef}$ and $D_{c0}$ coincide for magnetically soft magnetite nanoparticles.

## II. Numerical simulation

Dynamics of the unit magnetization vector $\vec{\alpha}(\vec{r})$ of a polycrystalline nanoparticle is described by the Landau – Lifshitz - Gilbert (LLG) equation [14,17]

$$\frac{\partial \vec{\alpha}}{\partial t} = -\gamma (\vec{\alpha} \times \vec{H}_{ef}) + \kappa \left( \vec{\alpha} \times \frac{\partial \vec{\alpha}}{\partial t} \right), \quad (1)$$

where $\gamma$ is the gyromagnetic ratio and $\kappa$ is the phenomenological damping constant. The effective magnetic field $\vec{H}_{ef}$ acting on the unit magnetization vector can be calculated as a derivative of the total nanoparticle energy $W$ [14]

$$\vec{H}_{ef} = -\frac{\partial W}{V M_s \partial \vec{\alpha}};$$

$$M_s \vec{H}_{ef} = C \Delta \vec{\alpha} - \frac{\partial w_a}{\partial \vec{\alpha}} + M_s \vec{H}'. \quad (2)$$

Here $V$ is the nanoparticle volume, $M_s$ is the saturation magnetization, $C = 2A$ is the exchange constant, $w_a(\vec{\alpha})$ is the magneto-crystalline anisotropy energy density, and $\vec{H}'$ is the demagnetizing field.

For numerical simulation a nanoparticle is approximated by a set of small ferromagnetic cubes of side $b$ much smaller than the exchange length, $L_{ex} = \sqrt{C}/M_s$, of the ferromagnetic material. Typically, several thousands of numerical cells, $N \sim 10^3 - 10^4$, are necessary to approximate with sufficient accuracy stationary magnetization distributions in nanoparticle volume. The equilibrium micromagnetic configurations in the nanoparticles studied were calculated starting from arbitrary initial micromagnetic state, the magnetic damping parameter being $\kappa = 0.5$. In accordance with the Eq. (1), the final magnetization state is assumed to be stable under the condition

$$\max_{(1 \leq i \leq N)} \left[ \left| \vec{\alpha}_i \times \vec{H}_{ef,i} \right| / \left| \vec{H}_{ef,i} \right| \right] < 10^{-6}, \quad (3)$$

where $\vec{\alpha}_i$ and $\vec{H}_{ef,i}$ are the unit magnetization vector and effective magnetic field in the $i$-th numerical cell, respectively.

To simulate a distribution of the easy anisotropy axes in a polycrystalline magnetic nanoparticle, we first select randomly within a particle volume several seed cells, $N_g$ = 4 - 16, which serve as the nuclei of the monocrystalline grains to be constructed. Then we consistently attach the closest numerical cells to every embryo until all available numerical cells are connected to one of the growing grains. Using such an algorithm one can generate within the nanoparticle volume the disjoint continuous areas representing the monocrystalline grains of a polycrystalline nanoparticle. Interestingly, this algorithm leads to the partition of the particle volume into $N_g$ grains of similar volume. Other partitioning algorithms that have been tried lead to similar results.

The directions of the easy anisotropy axes in the crystallites created in such a manner were selected randomly. For a nanoparticle with uniaxial magnetic anisotropy the magneto-crystalline anisotropy energy density of the $j$-th monocrystalline grain is given by

$$w_{a,j} = K \left( 1 - (\vec{\alpha} \vec{n}_j)^2 \right), \quad j = 1,2,..N_g, \quad (4a)$$

where $K$ is the uniaxial anisotropy constant and $\bm{n}_j$ is the unit vector parallel to the easy anisotropy axis of the given grain. For a nanoparticle with cubic anisotropy the corresponding expression reads

$$w_{a,j} = K_c \left( (\vec{\alpha} \vec{e}_{1j})^2 (\vec{\alpha} \vec{e}_{2j})^2 + (\vec{\alpha} \vec{e}_{1j})^2 (\vec{\alpha} \vec{e}_{3j})^2 + (\vec{\alpha} \vec{e}_{2j})^2 (\vec{\alpha} \vec{e}_{3j})^2 \right)$$
(4b)

Here $K_c$ is the cubic magnetic anisotropy constant, and ($\bm{e}_{1j}$, $\bm{e}_{2j}$, $\bm{e}_{3j}$) is a set of orthogonal unit vectors that determine an orientation of $j$-th monocrystalline grain of the nanoparticle. One may hope that this numerical model is capable to describe the distribution of the easy anisotropy axes in real polycrystalline nanoparticles created as a result of the coalescence of monocrystalline embryos originally formed by a chemical reaction in a solution [6].

Fig. 1 shows typical examples of a random partitioning of a quasi-spherical nanoparticle into different numbers $N_g$ = 4, 8, 16 monocrystalline granules of approximately equal volume obtained by the algorithm described above. It is assumed that the type of the magnetic anisotropy in all granules of the given magnetic nanoparticle is the same. However, as we mentioned above the easy anisotropy axes are randomly oriented in different granules of a polycrystalline nanoparticle.

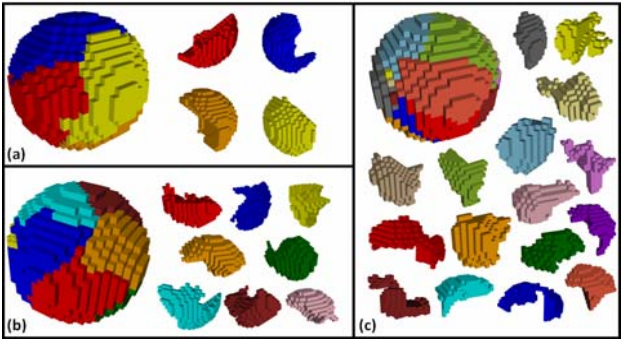

Fig. 1. Examples of a random partition of a spherical nanoparticle into 4 (a), 8 (b) and 16 (c) single crystal grains, respectively.

When a subdivision of a particle of a given diameter into monocrystalline granules is carried out, and the directions of the easy anisotropy axes are assigned to each numerical cell, the stationary magnetization distributions existing in a given nanoparticle can be obtained by solving the dynamic LLG Eq. (1). The calculations show that for a given polycrystalline nanoparticle in the range of diameters studied only one or two stable stationary micromagnetic states usually exist. They are a quasi-uniform state with average magnetization close to the saturation magnetization, and a vortex with relatively small remanent magnetization. In this paper, for each nanoparticle of the given type and geometry, a sufficiently large number of independent random particle partitions with the fixed number $N_g$ of monocrystalline granules was generated. Then, for this assembly of polycrystalline nanoparticles the average energy and average magnetization of low-lying micromagnetic states were calculated.

Table. 1. Magnetic parameters of the nanoparticles studied [26,27].

|  | $M_s$, emu/cm$^3$ | $A$, erg/cm | $K$, erg/cm$^3$ | $L_{ex}$, nm |
|---|---|---|---|---|
| Co | 1400 | $1.3\times10^{-6}$ | $4.3\times10^6$ | 11.5 |
| CrO$_2$ | 490 | $4.37\times10^{-7}$ | $3.0\times10^5$ | 19.1 |
| CoFe$_2$O$_4$ | 420 | $1.5\times10^{-6}$ | $2.0\times10^6$ | 41.2 |
| Fe$_3$O$_4$ | 480 | $1.0\times10^{-6}$ | $-1.0\times10^5$ | 29.5 |

The material parameters of various types of the nanoparticles studied are given in Table 1. In order to investigate the effect of complicated crystal structure on the particle magnetic properties, the nanoparticles of both uniaxial (Co, CrO$_2$) and cubic (CoFe$_2$O$_4$, Fe$_3$O$_4$) types of magnetic anisotropy were considered.

## III. Results and discussion

Let us consider the results of numerical simulation obtained for polycrystalline magnetic nanoparticles listed in Table 1, in comparison with the properties of monocrystalline nanoparticles of the same composition.

**Cobalt nanoparticles**

The magnetic properties of spherical polycrystalline cobalt nanoparticles have been studied in Ref. 25. Meanwhile, in the experiment cobalt nanoparticles of various shapes can be found. Therefore, in this paper oblate and elongated spheroidal cobalt nanoparticles with aspect ratios $D_z/D$ = 2/3 and 3/2, respectively, were also studied for completeness. Here $D_z$ and $D$ are the longitudinal and transverse diameters of a spheroid, respectively. It is assumed that monocrystalline cobalt granules have a hexagonal crystal structure, so that the energy density of magnetic anisotropy of individual cobalt granules is given by Eq. (4a) with the corresponding uniaxial anisotropy constant, $K = 4.3\times10^6$ erg/cm$^3$.

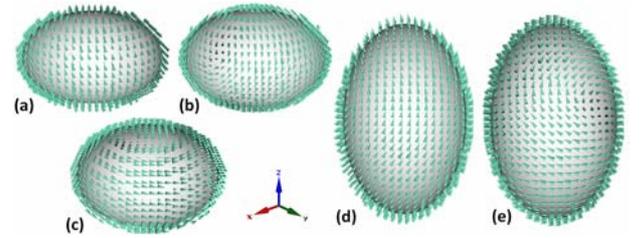

Fig. 2. Stationary magnetization distributions in polycrystalline cobalt nanoparticles of different diameters with various aspect ratios: a), b), c) $D_z/D$ = 2/3; d, e) $D_z/D$ = 1.5.

Figs. 2a - 2c show the calculated numerically stationary magnetization distributions in polycrystalline cobalt nanoparticles with the aspect ratio $D_z/D$ = 2/3. Fig. 2a shows a quasi-uniform state in a particle with a transverse diameter $D$ = 24 nm. Figs. 2b and 2c show vortex states with different directions of the vortex axis in particles with diameters $D$ = 56 nm, and $D$ = 44 nm, respectively. Figs. 2d and 2e show the quasi-uniform and vortex states in cobalt nanoparticles with the aspect ratio $D_z/D$ = 1.5 and with transverse diameters $D$ = 20 nm, and $D$ = 32 nm, respectively. It is worth of mentioning that the structures of the quasi-uniform and vortex states in polycrystalline nanoparticles are close to analogous magnetization distributions in monocrystalline cobalt particles. However, in polycrystalline particles the directions of the average magnetization of the quasi-uniform state, as well as the vortex axis directions depend on the specific distribution of monocrystalline grains over the nanoparticle volume. In addition, the total energies of the stationary states for polycrystalline particles can differ significantly from that for monocrystalline particles of the same geometry.

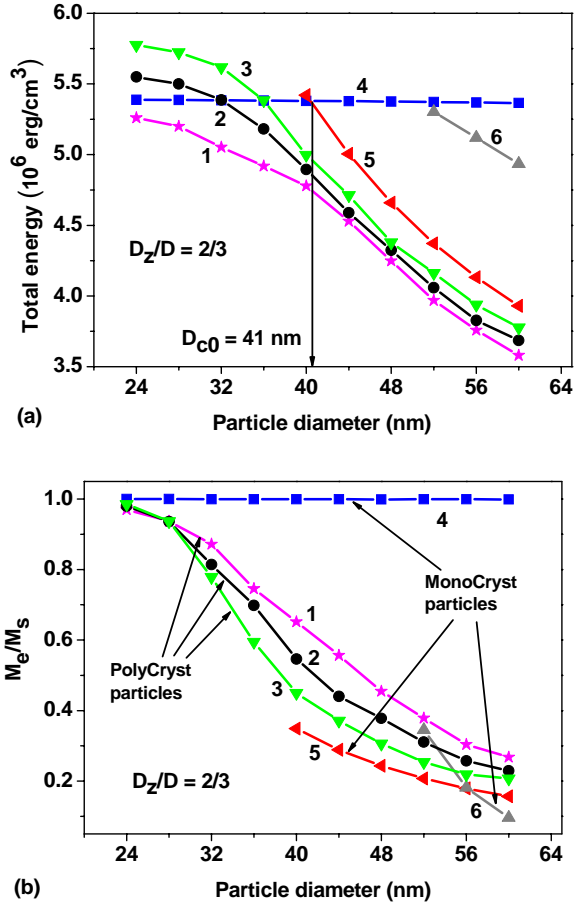

Fig. 3. a) Energy diagram of stationary micromagnetic states in oblate cobalt nanoparticles; b) the total reduced magnetic moment of various states as a function of the transverse diameter $D$. Curves 1-3 and curves 4 to 6 correspond to polycrystalline and single crystal particles, respectively.

Fig. 3a shows the energy diagram of stationary micromagnetic states in monocrystalline and polycrystalline cobalt nanoparticles with the aspect ratio $D_z/D = 2/3$. Curves 4 - 6 in Fig. 3a correspond to the case of a monocrystalline cobalt nanoparticle with easy anisotropy axis parallel to the short diameter of the spheroid, $D_z$. In this nanoparticle for all investigated diameters there is a stable uniform state with a magnetization parallel to the easy anisotropy axis (curve 4). The vortex state with the vortex axis oriented perpendicular to the easy anisotropy axis competes in energy with the uniform magnetization, (curve 5 in Fig. 3a). Curve 6 in Fig. 3a shows the energy of the vortex whose axis is oriented parallel to the easy anisotropy axis. The intersection of the curves 4 and 5 in Fig. 3a determines the single-domain diameter, $D_{c0} = 41$ nm, of the oblate monocrystalline cobalt nanoparticle with aspect ratio $D_z/D = 2/3$. For the oblate particle the transverse vortex remains stable in the range of diameters $D \geq 40$ nm.

Curves 1-3 in Fig. 3a show the average energies of stationary micromagnetic states in assemblies of polycrystalline oblate nanoparticles with different amounts of crystalline granules, namely, $N_g = 4$ for curve 1, $N_g = 8$ for curve 2, and $N_g = 16$ for curve 3, respectively. To obtain statistically reliable results, curves 1-3 were obtained by averaging over 200 – 250 independent realizations of polycrystalline oblate nanoparticles of fixed diameter in the range of sizes 24 nm $\leq D \leq 60$ nm. The dependence of the average reduced magnetic moment of oblate cobalt nanoparticles on the transverse particle diameter is shown in Fig. 3b. Curves 4 and 5 in Fig. 3b give the magnetization of a monocrystalline cobalt nanoparticle in uniform and transverse vortex states, respectively, Curves 1- 3 correspond to polycrystalline nanoparticles with various number of granules $N_g = 4$, 8 and 16, correspondingly.

Comparing the curves 1-3 in Fig. 3a and 3b, one can conclude that the stationary states in polycrystalline particles with the smallest number of granules, $N_g = 4$, (curves 1), have the smallest average energy and, at the same time, the largest average magnetic moment. With an increase in the number of granules in the polycrystalline nanoparticle, the average particle energy increases, whereas average magnetic moment decreases. This effect is a direct consequence of a subdivision of the nanoparticle volume into the domains with different directions of the easy anisotropy axes.

The calculations show that only quasi- uniform states with a reduced magnetic moment $M_e/M_s > 0.9$ are realized for oblate polycrystalline cobalt nanoparticles with diameters $D \leq 28$ nm, and only vortex states are realized in the range of diameters $D \geq 52$ nm. In the intermediate range of sizes, 28 nm $< D <$ 52 nm, both vortex and quasi- uniform micromagnetic states can be found in oblate polycrystalline nanoparticles. Based on the results obtained, one can conclude that in oblate polycrystalline cobalt nanoparticles quasi- uniform micromagnetic states arise in the interval of transverse diameters $D \leq 28$ nm, regardless of the number of monocrystalline granules in the particle. Accordingly, the value $D_{c,ef} = 28$ nm can be taken as the effective single-domain diameter of the polycrystalline cobalt nanoparticle with aspect ratio $D_z/D = 2/3$. The latter is thus much smaller than the single-domain diameter, $D_{c0} = 41$ nm, of the oblate monocrystalline cobalt nanoparticle.

Fig. 4 shows the probability of the appearance of stationary micromagnetic states with different average magnetization, $M_e/M_s$, for oblate polycrystalline cobalt nanoparticles in the intermediate range of diameters. As Fig. 4a demonstrates, for particles with diameter $D$ = 32 nm, close to the effective single-domain diameter $D_{c,ef}$ = 28 nm, a significant fraction of the stationary states still have large average magnetic moments, $M_e/M_s \geq 0.8$. For nanoparticles with diameter $D$ = 40 nm (see Fig. 4b), near the middle of the intermediate range of diameters, the probability of appearance of the vortex states with reduced remanent magnetization $M_e/M_s$ = 0.3 – 0.5 is nearly the same as for quasi uniform states. Finally, as Fig. 4c shows, for nanoparticles with diameter $D$ = 48 nm the majority of the stationary micromagnetic states are vortices with reduced remanent magnetization $M_e/M_s$ = 0.2 – 0.4. These probabilities are calculated for an assembly of 200 polycrystalline nanoparticles consisting of $N_g$ = 4 monocrystalline grains for each transverse diameter studied. Similar results are obtained also for assemblies of oblate polycrystalline cobalt nanoparticles consisting of $N_g$ = 8 and 16 monocrystalline grains, respectively.

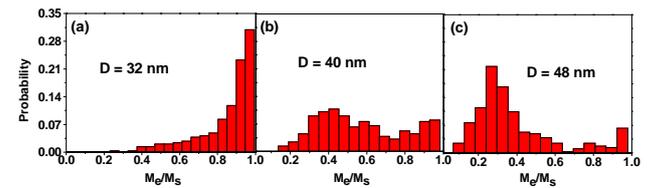

Fig. 4. The probability of the appearance of stationary micromagnetic states with different mean magnetization in polycrystalline oblate cobalt nanoparticles of various transverse diameters.

Fig. 5a shows the energy diagram of stationary micromagnetic states of elongated monocrystalline and polycrystalline cobalt nanoparticles with aspect ratio $D_z/D$ = 1.5. Fig. 5b shows the total reduced magnetic moment of these nanoparticles as a function of transverse diameter $D$. Curves 1-3 correspond to polycrystalline elongated nanoparticles with different amounts of single crystal grains. Curve 4 corresponds to a uniform state in a monocrystalline cobalt nanoparticle with an easy anisotropy axis parallel to particle symmetry axis. Curve 5 corresponds to the vortex with the vortex axis parallel to the easy anisotropy axis. As can be seen in Fig. 5a, the single-domain diameter of an elongated monocrystalline cobalt nanoparticle with aspect ratio $D_z/D$ = 1.5 equals $D_{c0}$ = 56 nm. The vortex state in this nanoparticle is stable at $D \geq 36$ nm.

Detailed calculations carried out for polycrystalline cobalt nanoparticles with aspect ratio $D_z/D$ = 1.5 in the range of transverse diameters 20 nm $\leq D \leq$ 64 nm showed that only quasi-uniform states with reduced magnetic moment close to unity exist in the interval $D \leq$ 24 nm, whereas in the range of transverse diameters $D \geq$ 44 nm only vortex states are realized. In the intermediate range of diameters, 24 nm $< D <$ 44 nm, in elongated polycrystalline cobalt nanoparticles both vortex and quasi uniform micromagnetic states can be found. Thus, the value $D_{c,ef}$ = 24 nm can be taken as the effective single-domain diameter of polycrystalline cobalt nanoparticle with aspect ratio $D_z/D$ = 1.5.

**Chromium dioxide**

Similar calculations have been made for spherical chromium dioxide nanoparticles with uniaxial type of magnetic anisotropy,

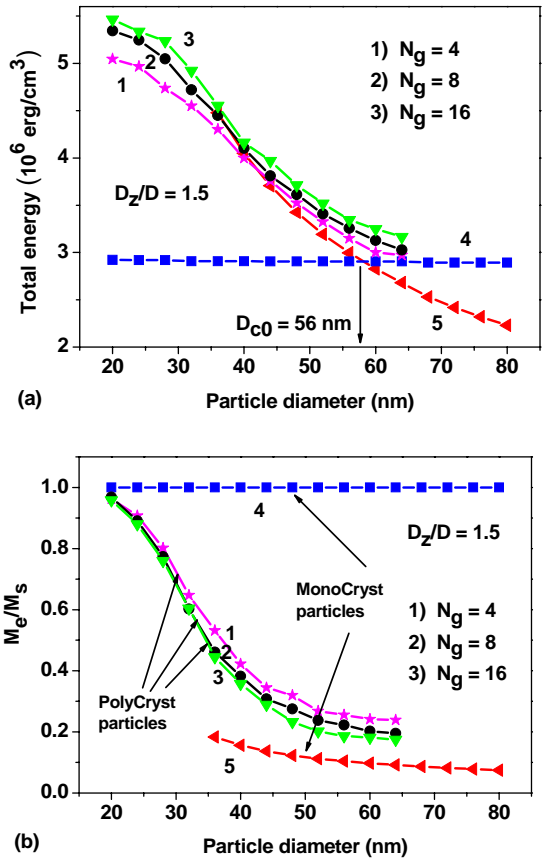

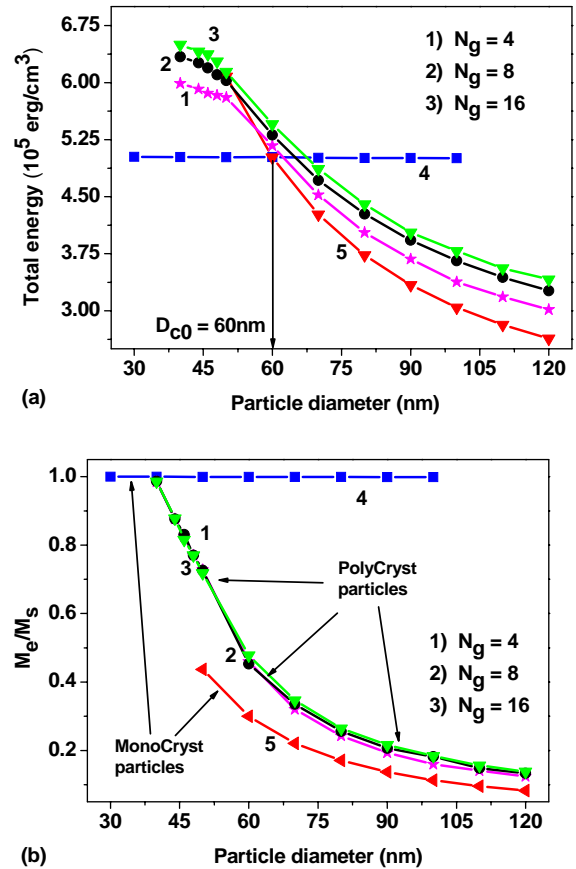

Fig. 5. a) Energy diagram of stationary micromagnetic states in elongated cobalt nanoparticles with aspect ratio $D_z/D$ = 1.5; b) the total reduced magnetic moment as a function of the transverse diameter $D$. Curves 1-3 and curves 4, 5 correspond to polycrystalline and single crystal particles, respectively.

Fig. 6. a) Energy diagram of stationary micromagnetic states in chromium dioxide nanoparticles; b) the total reduced magnetic moment as a function of particle diameter. Curves 1-3 correspond to polycrystalline nanoparticles, and curves 4, 5 to monocrystalline ones, respectively.

but much lower saturation magnetization than that of cobalt.

As can be seen in Fig. 6a, for a monocrystalline particle of chromium dioxide the single-domain diameter is given by $D_{c0} = 60$ nm. The calculations of stationary magnetization distributions for polycrystalline chromium dioxide nanoparticles were carried out in the interval of diameters 40 nm ≤ $D$ ≤ 120 nm. It was found that only quasi-uniform states are realized in the range of diameters $D$ ≤ 44 nm, and only vortices exist for $D$ ≥ 60 nm. In the transition region, 44 nm < $D$ < 60 nm, both vortices and quasi uniform magnetization distributions are realized with various probabilities. Based on these calculations, the effective single-domain diameter of polycrystalline chromium dioxide nanoparticles was determined to be $D_{c,ef} = 44$ nm.

**Cobalt ferrite**

To investigate the effect of the type of magnetic anisotropy on the magnetic properties of polycrystalline nanoparticles, spherical nanoparticles of cobalt ferrite, with positive cubic anisotropy constant, $K_c = 2.0 \times 10^6$ erg/cm$^3$, were considered.

Consequently, monocrystalline cobalt ferrite granules have 6 equivalent directions of the easy anisotropy axis. The energy density of magnetic anisotropy of individual cobalt ferrite grain is given by Eq. (4b).

As Fig. 7a shows, for a monocrystalline cobalt ferrite nanoparticle the single-domain diameter is given by $D_{c0} = 140$ nm. The calculations of stationary magnetization distributions for polycrystalline cobalt ferrite nanoparticles were carried out in the interval of diameters 60 nm ≤ $D$ ≤ 150 nm. In Figs. 7b - 7e the probabilities of the appearance of stationary micromagnetic states with different mean magnetization are shown for polycrystalline cobalt ferrite nanoparticles of various diameters consisting of $N_g = 4$ monocrystalline grains. One can see that for polycrystalline cobalt ferrite nanoparticles with diameter $D = 110$ nm most of the stationary micromagnetic states have large average magnetization, $M_e/M_s > 0.9$. The same is true for nanoparticles with diameters $D < 110$ nm. On the other hand, for nanoparticles with diameter $D \geq 120$ the probability of appearance of non-uniform micromagnetic state with average magnetization, $M_e/M_s \sim 0.5$, increases gradually as a function of the particle diameter. One can conclude safely, that the lower bound for the effective single-domain diameter of polycrystalline cobalt ferrite nanoparticles is given by $D_{c,ef} = 110$ nm.

**Magnetite**

The magnetic properties of polycrystalline magnetite nanoparticles are particularly interesting, since magnetite nanoparticles are widely used in biomedicine [1-3]. Magnetite nanoparticles also have cubic type of magnetic anisotropy. But unlike cobalt ferrite, the cubic anisotropy constant for magnetite is negative, $K_c = -1.0 \times 10^5$ erg/cm$^3$. As a result, monocrystalline magnetite granules have 8 equivalent directions of the easy anisotropy axis. Fig. 8 shows the energy of stationary micromagnetic states calculated for spherical monocrystalline and polycrystalline magnetite nanoparticles with different amounts of crystalline granules in the range of diameters 32 nm ≤ $D$ ≤ 96 nm.

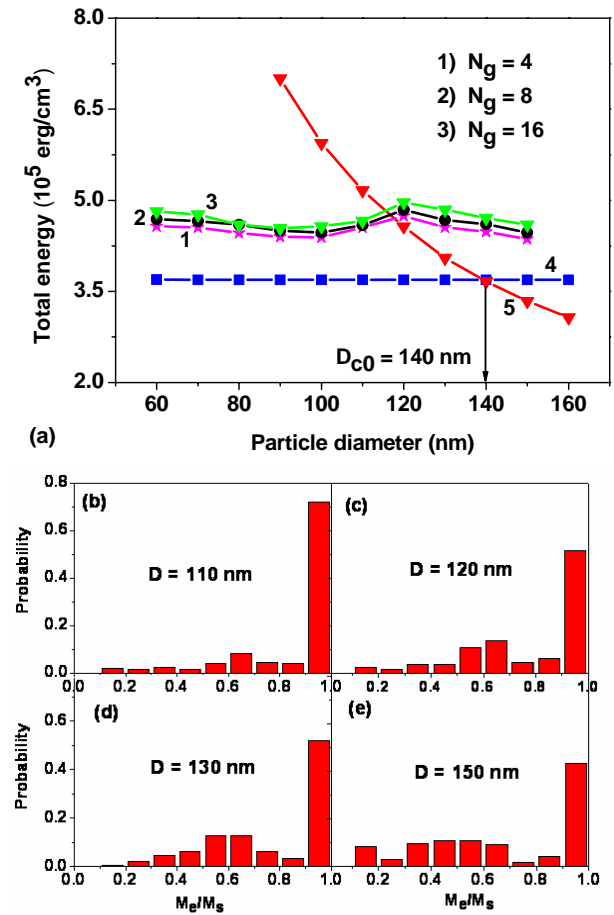

Fig. 7. a) Energy diagram of stationary micromagnetic states in monocrystalline and polycrystalline cobalt ferrite nanoparticles; b) – e) probability of the appearance of stationary micromagnetic states with different mean magnetization in polycrystalline cobalt ferrite nanoparticles of various transverse diameters.

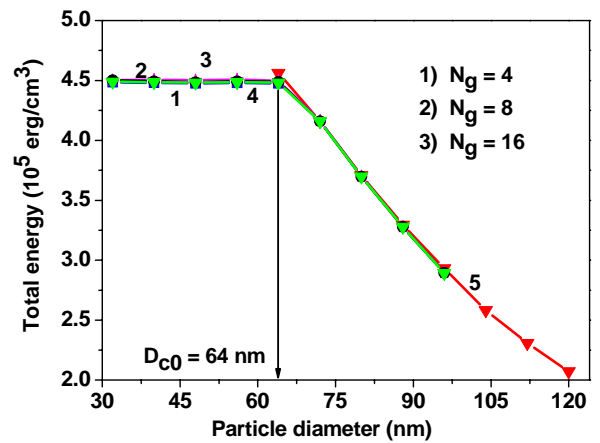

Fig. 8. Energy diagram of stationary micromagnetic states in spherical polycrystalline (curves 1-3) and monocrystalline (curves 4, 5) magnetite nanoparticles.

As can be seen in this figure, the effect of complicated crystal structure on the properties of magnetite nanoparticles is insignificant. In particular, single-domain diameters for monocrystalline and polycrystalline magnetite nanoparticles coincide, $D_{c0} = D_{c,ef} = 64$ nm.

In Table 2 we present the calculated single-domain diameters, $D_{c0}$, of the monocrystalline nanoparticles studied in comparison with the effective single-domain diameters of the polycrystalline ones, $D_{c,ef}$. In addition, the values of the parameter $p = N_z M_s^2 / 2K$, which characterizes the magnetic hardness of the nanoparticles, are also given in Table 2. Note, that one has $p >> 1$ and $p < 1$ for particles of soft and hard magnetic types, respectively. First of all, one can see that a noticeable difference between the $D_{c0}$ and $D_{c,ef}$ values exists for magnetic nanoparticles with uniaxial type of magnetic anisotropy. Furthermore, this difference increases with increasing of magnetic hardness of the nanoparticle. The greatest difference between the $D_{c0}$ and $D_{c,ef}$ diameters was obtained for elongated cobalt nanoparticles with $p = 0.667$. On the other hand, for nanoparticles with cubic anisotropy, relatively small difference in $D_{c0}$ and $D_{c,ef}$ diameters exists only for cobalt ferrite nanoparticles with a small parameter $p = 0.185$. For magnetically soft magnetite nanoparticles the single-domain diameters $D_{c0}$ and $D_{c,ef}$ coincide. Obviously, the individual monocrystalline grains with cubic anisotropy have a large number of equivalent directions of the easy anisotropy axis. Therefore, for polycrystalline nanoparticle with cubic anisotorpy the probability to get an inhomogeneous distribution of easy anisotropy axes over the particle volume is much smaller than for particles with uniaxial type of magnetic anisotropy. As a result, some difference between the $D_{c0}$ and $D_{c,ef}$ values for polycrystalline nanoparticle with cubic anisotropy appears only for small values of parameter $p$. This gives a qualitative physical explanation for the results of the numerical simulation presented in Table 2.

Table. 2. The calculated single-domain diameters for mono and polycrystalline magnetic nanoparticles.

|   | $D_z/D$ | $p$ | $D_{c0}$, nm | $D_{c,ef}$, nm | Anisotropy type |
|---|---|---|---|---|---|
| Co | 2/3 | 1.277 | 41 | 28 | Uniaxial |
| Co | 1 | 0.955 | 45[25] | 24[25] | Uniaxial |
| Co | 1.5 | 0.667 | 56 | 24 | Uniaxial |
| $CrO_2$ | 1 | 1.676 | 60 | 44 | Uniaxial |
| $CoFe_2O_4$ | 1 | 0.185 | 140 | 110 | Cubic |
| $Fe_3O_4$ | 1 | 4.825 | 64 | 64 | Cubic |

## IV. Conclusions

In this paper the influence of the crystal structure heterogeneities on the equilibrium magnetic properties of cobalt nanoparticles with different aspect ratio and spherical nanoparticles of chromium dioxide, cobalt ferrite and magnetite has been studied using numerical simulation. The nanoparticles studied possess both uniaxial and cubic types of magnetic anisotropy and have different magnetic hardness. For monocrystalline nanoparticles of these types the single-domain diameters $D_{c0}$ have been determined comparing the total energies of quasi-uniform and vortex micromagnetic states. The polycrystalline nanoparticles are modeled by means of subdivision of the nanoparticle volume into tightly bound single-crystal granules with randomly distributed directions of the easy anisotropy axes. The probability of appearance of quasi uniform and vortex states in sufficiently large assemblies of polycrystalline nanoparticles of various types have been calculated depending on the nanoparticle diameter. For polycrystalline nanoparticles the effective single-domain diameters, $D_{c,ef}$, have been determined under the condition that only quasi-uniform states with reduced magnetic moment $M_e/M_s > 0.9$ exist in the polycrystalline nanoparticle of a given type in the range of diameters $D < D_{c,ef}$.

It is shown that the subdivision of a nanoparticle into single-crystal granules with different orientations of the easy anisotropy axes substantially reduces the effective single-domain diameters for particles with uniaxial type of anisotropy. Moreover, for these nanoparticles the difference between the diameters $D_{c0}$ and $D_{c,ef}$ increases with increasing of the particle magnetic hardness. At the same time, for particles with cubic type of magnetic anisotropy, the influence of the crystal structure inhomogeneities on the equilibrium properties of the particles is not so important even for magnetically hard cobalt ferrite nanoparticles. It is practically absent for magnetically soft magnetite nanoparticles. This is a consequence of the fact that in particles with cubic type of magnetic anisotropy the probability to create an inhomogeneous easy axis distribution over the nanoparticle volume is small due to the presence of a large number of equivalent directions of the easy anisotropy axes in the individual granules.

## Acknowledgments

We acknowledge funding from Russian Ministry of Education and Science (Grant RFMEFI57815X0128).